# On the group-theoretical approach to the study of interpenetrating nets[1]

Igor A. Baburin

February 15th, 2016

*Theoretische Chemie, Technische Universität Dresden, Bergstraße 66b, 01062 Dresden*

*Correspondence email: baburinssu@gmail.com*

**Synopsis** A general theoretical framework based on group–subgroup and group–supergroup relations is proposed to describe and to derive interpenetrating nets.

**Abstract** Using group–subgroup and group–supergroup relations, we develop a general theoretical framework to describe and to derive interpenetrating three-periodic nets. The generation of interpenetration patterns is readily accomplished by replicating a single net with a supergroup $G$ of its space group $H$ under the condition that site symmetries of vertices and edges are the same in both $H$ and $G$. We show that interpenetrating nets cannot be mapped onto each other by mirror reflections because otherwise edge crossings would necessarily occur in the embedding. For the same reason any other rotation or roto-inversion axes from $G \setminus H$ are not allowed to intersect vertices or edges of the nets. This property significantly narrows the set of supergroups to be included in the derivation of interpenetrating nets. We describe a procedure based on the automorphism group of a *Hopf ring net* [Alexandrov, E. V., Blatov, V. A. & Proserpio, D. M. (2012). *Acta Cryst.* A**68**, 484–493] to determine maximal symmetries compatible with interpenetration patterns. The proposed approach is illustrated by working out examples of two-fold interpenetrated **utp**, **dia** and **pcu**, as well as multiple copies of enantiomorphic quartz (**qtz**) nets. Some applications to polycatenated 2-periodic layers are also discussed.

## 1. Introduction

One of the necessary symmetry conditions for a triply periodic *balance* surface to be free of self-intersections consists in that mirror reflections cannot map its two *labyrinth graphs* onto each other (Fischer & Koch, 1987). If a mirror plane exchanged the two labyrinths, then it would have to be embedded within the surface making it self-intersecting. If a self-intersection-free surface separates the two labyrinths, they are free of mutual edge crossings as well. Koch and Fischer (1988) tabulated symmetry conditions, namely, group–subgroup pairs which are compatible with triply periodic *balance* surfaces. These conditions are also fulfilled by pairs of interpenetrating, congruent three-periodic nets without edge crossings independently of whether a *balance* surface separating them from each other actually exists. However, the conditions of Fischer and Koch are in general too restrictive because interpenetrating nets can be separated by self-intersecting surfaces (Koch, 2000*a*) or if three or more nets interpenetrate. In this paper we generalize the result of Fischer and Koch and show that mirror reflections acting between any interpenetrating three-periodic nets would necessarily enforce edge crossings in a Euclidean embedding. This observation can be extended to any rotation or rotoinversion axis mapping the nets onto each other and at the same time intersecting their vertices and/or edges. Based on group–supergroup relations, we propose a general group-theoretical approach to construct interpenetrating nets. Additionally, to find maximal symmetries of interpenetration patterns, we introduce a systematic procedure that is based on the automorphism group of a *Hopf ring net* (Alexandrov *et al.*, 2012) and makes use of group–subgroup as well as group–supergroup relations. Throughout the paper we illustrate our approach by examples.

---

[1] Some ideas of this paper were presented by the author in his talk at the conference of the Italian Crystallographic Association (AIC) in Florence, September 18th, 2014.

## 2. Preliminaries and results

We consider a set $\Gamma$ of *n* symmetry-related interpenetrating three-periodic nets $\Gamma_i$, *i*=1, 2…*n* (for a definition of a *periodic net* see *e.g.* Delgado-Friedrichs and O'Keeffe, 2005) in a Euclidean embedding with a three-dimensional space group *G*. Accordingly, the vertices of nets are just points in the 3D Euclidean space and the edges are straight line segments. We assume no edge crossings in the embedding. A group *G* acts transitively on a set $\Gamma$, *i.e.* it maps the nets onto themselves as well as onto each other. The elements of *G* which fix (as a whole) an arbitrarily chosen net $\Gamma_i$ generate the *stabilizer* of this net in *G* denoted as $G_{\Gamma_i}$ [for the action of groups on graphs consult *e.g.* Beineke, Wilson, Cameron, 2004]. Let *H* be a restriction of $G_{\Gamma_i}$ to $\Gamma_i$. Since the group *H* is actually isomorphic to $G_{\Gamma_i}$, the index of *H* in *G* equals *n* (*cf.* Koch *et al.*, 2006). As a consequence, we immediately obtain the following lemma.

*Lemma ('on stabilizers')*. Let $\Gamma$ be a set of *n* symmetry-equivalent three-periodic nets $\Gamma_i$ (*i*=1, 2…*n*) with a three-dimensional space group *G* of the whole set. The elements of *G* which map a net $\Gamma_i$ onto itself form a group *H*. Then vertex and edge stabilizers (=*site-symmetry groups*) of $\Gamma_i$ in *H* are isomorphic to those in the group *G*.

*Remark 1*. Strictly speaking, if subgroup *H* is not normal in *G*, then the nets $\Gamma_i$ correspond to different subgroups $H_i$ (*i*=1, 2,…*k*, *k* ≤ *n*) which are conjugate in *G* (Koch *et al.*, 2006). We would not distinguish between conjugate subgroups $H_i$ in the following because this is not important in the context of our results.

The lemma 'on stabilizers' is crucial for the proof of the theorem below.

*Theorem*. Let $\Gamma$ be a set of *n* symmetry-equivalent interpenetrating three-periodic nets without edge crossings in a Euclidean embedding with a three-dimensional space group *G*. The elements of *G* which map a net $\Gamma_i$ onto itself generate its subgroup *H*. Then the cosets of *H* in *G* do not contain mirror reflections.

*Proof*. We proceed by *reductio ad absurdum*. Let *G* contain mirror reflections additional to those in the subgroup *H*. Consider a Euclidean embedding of a net $\Gamma_i$ with a space group *H*. Let us describe the action of a mirror plane (arbitrarily oriented in the unit cell of *H*) on $\Gamma_i$. Since $\Gamma_i$ is a connected graph, a mirror plane would necessarily intersect some of its vertices and/or edges. In fact, there are four different possibilities (which may occur in combination):

*(a)* a mirror plane passes through some vertices of $\Gamma_i$. As a result, vertex stabilizers of $\Gamma_i$ in *G* would be enhanced compared with those in *H*, a contradiction with the above lemma;

*(b)* a mirror plane perpendicularly bisects some edges of $\Gamma_i$. In this case edge stabilizers of $\Gamma_i$ in *G* would be enhanced, once again contrary to the above lemma;

*(c)* a mirror plane runs perpendicularly to some edges of $\Gamma_i$ but does not go through their midpoints. As a consequence, a mirror plane would generate another net, say, $\Gamma_j$ from the original one. However, the edges of $\Gamma_j$ will be collinear with their preimages from $\Gamma_i$ and, furthermore, will be partially coincident. We qualify this case as a special kind of edge crossing;

*(d)* a mirror plane intersects some edges of $\Gamma_i$ which are arbitrarily inclined with respect to it. This situation corresponds to a usual case of edge crossing (Fig. 1).

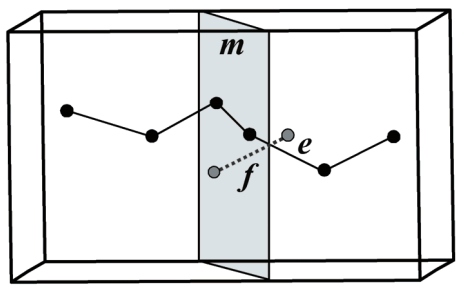

**Figure 1**. An arbitrary edge path in the net $\Gamma_i$. A mirror plane *m* runs through the edge *e* that intersects its image *f*.

In conclusion, a mirror plane of *G* (not contained in *H*) either enhances vertex and/or edge stabilizers of $\Gamma_i$ in *G* that is forbidden by the lemma or otherwise induces edge crossings. This means that the nets in a set Γ cannot be mapped onto each other by mirror reflections[2] or, in other words, mirrors cannot belong to the cosets of *H* in *G*. □

In fact, the theorem and the preceding lemma have a wider applicability. They hold if *G* and *H* are arbitrary isometry groups in 3D Euclidean space (point groups, rod groups, layer or space groups), if *G* acts transitively on the objects (*e.g.* polyhedra, helices, layers *etc*.), each with a group *H* that has a finite index in *G*. This means that intergrowths of symmetry-related polyhedra[3], interwoven helices (or more generally, braids with all strands equivalent) and interpenetrated layers (for a recent review see Carlucci *et al*., 2014) are covered by the theorem. An extension to *polycatenated* layers requires more specification because in this case the index of *H* (that is a layer group) in a space group *G* is infinite. Namely, mirrors never map layers with different orientations onto each other (*inclined* polycatenation) while in the situation of *parallel* polycatenation mirrors cannot interchange layers with the same orientation. In the supporting information the reader can also find an intergrowth of three orthorhombically distorted tetrahedra (point group 222) which are related by a 3-fold rotation axis and therefore generate an arrangement with overall 23 symmetry.

---

[2] In terms of *interpenetration symmetry elements* introduced by Blatov *et al*. (2004) a mirror plane cannot be either a full interpenetration symmetry element (FISE) or a partial interpenetration symmetry element (PISE) so it cannot occur in any *interpenetration class*.
[3] However, edge crossings are often allowed for intergrowths of polyhedra so our restriction may be irrelevant.

*Remark 2.* Edge crossings caused by mirrors cannot be eliminated if edges would be replaced by arbitrary curved segments. If mirrors are retained, crossings are necessarily unavoidable. There exists a relation to self-intersecting minimal surfaces forming three-periodic labyrinths. Indeed, all such surfaces derived by Koch (2000*a*) do not admit mirror planes interchanging the labyrinths (corresponding group–subgroup pairs $G – H$ are $Pn\bar{3}m – Fd\bar{3}m$, $P6/mcc – P\bar{6}c2$, $P6/mcc – P6/m$, $P4/mcc – P4/mnc$, $P4/mcc – P4/m$, $Pccm – Cccm$, $Pccm – P112/m$, *i.e.* the theorem is fulfilled).

*Remark 3.* The theorem can be easily generalized to show that any symmetry element of finite order that is contained in $G \setminus H$ and intersects vertices and/or edges of $\Gamma_i$ either increases vertex or edge stabilizers in $G$ compared to $H$ (*e.g.* rotation or rotoinversion axes passing through the vertices or running entirely through the edges, two-fold axes bisecting perpendicularly the edges or inversion centers at their midpoints) – the situation forbidden by the lemma – or otherwise enforces edge crossings in the embedding. This observation points towards a certain pictorial property of interpenetrating nets: they *envelope* the scaffoldings of finite order symmetry elements belonging to the cosets of $H$ in $G$.

Furthermore, the theorem could lend further support for overwhelming occurrence of space groups without mirror planes which are usually considered to be unfavorable due to packing arguments (*e.g.* Vainshtein, 1982). An example would help at this point. From the results of Koch and Fischer (1988) it follows that a symmorphic space group $P2/m$ admits arrangements of 2-fold interpenetrated nets which can be described by only two group–subgroup pairs $G – H$: $P2/m – Pm$ and $P2/m – P2/m$ (2**a**) whereas 'analogous' asymmorphic group $P2/c$ allows in principle for arrangements corresponding to seven different group–subgroup pairs [$P2/c – P2$, $P2/c – P\bar{1}$, $P2/c – Pc$, $P2/c – P2_1/c$ (2**b**), $P2/c – C2/c$ (2**a**, 2**b**), $P2/c – P2/c$ (2**b**), $P2/c – P2/c$ (2**a**)]. Qualitatively speaking, asymmorphic space groups could accommodate a richer variety of patterns. As a simple consequence of the theorem, there cannot exist a three-dimensional racemate where *e.g.* two *chiral* nets are related by mirror planes. Thus, the role of asymmorphic centrosymmetric space groups is strengthened once again because they could be expected to preferentially occur as symmetry groups of achiral interpenetration patterns in general.

## 3. Applications: how to use the *group–supergroup approach* to generate interpenetration patterns and how to find their highest symmetry embeddings

### 3.1. Generation of interpenetrating nets

In the preceding sections we dealt with group–subgroup relations between the symmetry group acting transitively on a set of nets, $G$, and the symmetry group mapping an individual net onto itself, $H$. Now let us invert the procedure. Given a net embedding with a symmetry group $H$, how can we construct $n$ interpenetrated copies of it? It turns out that both the lemma and the theorem presented above can be effectively used to build up structure models of interpenetrating nets. To this end, we should first enumerate symmetry groups possible for the *embeddings*[4] of a certain (single) net. This could be done without principal difficulties by subgroup degradation (*International Tables for Crystallography*, V. A1, 2004) starting from the automorphism group (or a space group of the highest symmetry Euclidean embedding) of a net (Delgado-Friedrichs & O'Keeffe, 2003; Eon, 2011). To generate *n* interpenetrating nets, we replicate an embedding of a single net $\Gamma_i$ defined in a space group *H* by its supergroup *G* with a (finite) index *n* provided that we performed the transformation of basis vectors and coordinates from a subgroup to a supergroup. Note that, according to the lemma (see § 2), site-symmetries of all vertices ('*atoms*') and edges ('*bonds*') of the net $\Gamma_i$ in *G* should not change compared to its original space group *H*. If (due to special coordinate parameters) site-symmetries are not kept in the supergroup, then the construction of interpenetration pattern is in general impossible and it might be useful to look for the embeddings of a single net with lower symmetry. Additionally, in line with the theorem and our Remark 3 (§ 2) all supergroups of *H* which contain mirror planes or other symmetry elements of finite order additional to those in the group *H* and intersecting vertices or edges of the original net $\Gamma_i$ need not be considered[5]. Care must be taken when a group and a supergroup belong to different crystal systems, in which case it is convenient to specialize the metric of a unit cell of a subgroup to match that of a supergroup. By taking into account all possible group–supergroup pairs we can systematically derive entanglements being either topologically or geometrically different (including presumably non-ambiently isotopic ones, see *e.g.* Castle *et al.*, 2011). In enumerating group–supergroup pairs, the knowledge on equivalent supergroups (Koch, 1984) can be useful to avoid duplication. Additionally, the construction of a *Hopf ring net* (HRN) could be applied to filter out isotopically distinct patterns (Alexandrov *et al.*, 2012). Note that the first attempt to use group–supergroup relations for structure modeling was made to study conformations of two-fold interpenetrated diamondoid zinc imidazolates (Baburin & Leoni, 2010) without an explicit formulation of the mathematical background.

### 3.2. Maximal symmetry embeddings of interpenetrating nets

As already explained, interpenetration patterns for a given net embedding can be systematically derived by using group–supergroup relations. To characterize symmetry properties of interpenetrating nets, it makes sense to distinguish between *the maximal symmetry of an interpenetration pattern* (Fischer and Koch, 1976; Koch *et al.*, 2006; Alexandrov *et al.*, 2012) on one hand, and *the maximal symmetry possible for a system of interpenetrating nets*, on the other. It is well known (Koch *et al.*, 2006; Alexandrov *et al.*, 2012) that different nets can share the same interpenetration pattern, *i.e.* rings in systems of different interpenetrating nets can be catenated in an analogous way. However, it is often advantageous to know both maximal symmetry space groups. High symmetry embeddings for some *important* interpenetrating nets (**pcu**, **dia**, **srs**, **qtz**, **bto**, **hcb**, **sql**)[6] have been analyzed in a recent survey by Bonneau and O'Keeffe (2015). Unfortunately, their paper lacks a systematic approach to the problem. Here we make an attempt to fill this gap.

---

[4] In contrast to the enumeration of symmetry groups, the task to find *all* possible embeddings of a net (up to affine equivalence or ambient isotopy) is hardly feasible. Moreover, a net can have essentially different embeddings with the same space-group symmetry (*cf*. Koch & Sowa, 2004). Self-catenation phenomenon causes additional difficulties (Hyde & Delgado-Friedrichs, 2011). However, when building up crystal structure models in practice, the conformation of a net is fairly strictly fixed by stereochemical constraints (*e.g.* Baburin & Leoni, 2010).
[5] Sometimes (*e.g.* if edge crossings are due to 2-fold rotation axes), it is helpful to introduce auxiliary bi-coordinated nodes along the edges (thus making them 'curved' on purpose) and to avoid intersections of this kind.

Let us first show in more detail how group–supergroup relations can be used to determine the highest symmetry compatible with interpenetrating nets in some special cases. As earlier, we consider a set $\Gamma$ of $n$ interpenetrating nets $\Gamma_i$ ($i=1,2…n$) with a space group $G$ of the whole set while its subgroup $H$ (of index $n$) maps a net $\Gamma_i$ onto itself. For maximal symmetry embeddings we shall use the notation $G_{max}$ and $H_{max}$ with the same meaning. Additionally, for simplicity we consider interpenetration patterns only for crystallographic nets, *i.e.* for those whose automorphism group Aut($\Gamma_i$) is isomorphic to a space group (Klee, 2004; Eon, 2005), although our approach can be applied to non-crystallographic nets as well. At the moment we discuss the following situations:

(i) if Aut($\Gamma_i$) is known, it is straightforward to check if the number of vertex orbits of $\Gamma_i$ in $G$ remains the same as in Aut($\Gamma_i$) and if the vertex stabilizers in $G$ are isomorphic to those in Aut($\Gamma_i$). This implies $H_{max}$ = Aut($\Gamma_i$). Once these conditions are met, then one can guarantee that maximal symmetry embedding has been found for a set $\Gamma$ (however, $G_{max}$ may not be unique, *cf.* § 4.1). *Example*: eight interpenetrating **srs** nets (**srs-c8**) with symmetry $G = I432$ (Hyde & Ramsden, 2000; Bonneau & O'Keeffe, 2015; see also Fig. 19 in Evans *et al.*, 2015). The vertices occupy the Wyckoff position $I432$ 8$c$ .32. In this case $H = I4_132$ (2**a**) coincides with the automorphism group of an **srs** net (vertices at $I4_132$ 8$a$ .32).

(ii) if $H \neq$ Aut($\Gamma_i$) that is most common, one has to adopt a more elaborate step-by-step procedure. *Example:* five-fold '*abnormal*' interpenetration[7] of diamond nets [$G = I4_1/a$, $H = I4_1/a$ (2**a**–**b**, **a**+2**b**, **c**)] with vertices at $I4_1/a$, 4$a$ ($\bar{4}$) (we propose a name **dia-c5\*** for it)[8]. It was first observed in the crystal structure of adamantane-1,3,5,7-tetracarboxylic acid (Ermer, 1988; Batten & Robson, 1998). To look for the maximal symmetry of this pattern, those supergroups $G'$ of $G$ need to be considered which (a) admit 'analogous' subgroups $H'$ of the same index as the index of $H$ in $G$ and where (b) site-symmetries of vertices are necessarily enhanced with respect to $G$. Out of minimal supergroups of $I4_1/a$ only the $P4_2/n$ group has a subgroup of index 5 that corresponds to cell enlargement in the ***a*,*b***-plane (*cf.* Müller, 1995). However, the enhancement of site symmetry is not possible because $P4_2/n$ belongs to the same crystal class as $I4_1/a$. Hence, the space group of this interpenetration pattern cannot be further increased, *i.e.* $I4_1/a$ represents its maximal symmetry.

---

[6] We use here bold three-letter symbols for nets as proposed by M. O'Keeffe (O'Keeffe *et al.*, 2008).

[7] For diamond nets a distinction between *normal vs. abnormal* modes of interpenetration is based on whether or not the nets interpenetrate along the (at least topological) $\bar{4}$ axis of the adamantane cage (*e.g.* Batten & Robson, 1998). For example, in the '*normal*' **dia-c5** pattern nets are related by a translation vector along the tetragonal *c* axis while in the '*abnormal*' entanglement **dia-c5\*** nets are displaced relative to each other by translations in ***a*** (or ***b***) tetragonal directions.

[8] An interpenetrating sphere packing $t[3/4/t1]^5$ (Koch *et al.*, 2006) shows the same interpenetration pattern.

Another more general and systematic way to find maximal symmetry for interpenetrating nets is to deal with the automorphism group of the respective *Hopf ring net* (HRN). Recall that the vertices of HRN are barycenters of catenated *strong rings* (for a definition see *e.g.* Delgado-Friedrichs and O'Keeffe, 2005) while the edges correspond to the Hopf links between them, *i.e.* this net characterizes catenation pattern up to ambient isotopy (Alexandrov *et al.*, 2012). The construction of HRN is implemented in ToposPro (Blatov *et al.*, 2014) which greatly facilitates the analysis of entanglements in crystal structures and model nets. Note that HRN is usually a relatively complicated graph with high valences of vertices, and therefore, difficult to visualize. The reader is referred to the paper by Alexandrov *et al.* (2012) for the drawings of HRNs for common interpenetrating nets. Here we shall use the HRN as a formal means to describe the symmetry of an interpenetration pattern. Note that if three-periodic nets interpenetrate, then HRN is very often a connected three-periodic graph. The connectedness of HRN is reasonable to conjecture especially if *all strong rings* of the nets are catenated. However, if some strong rings (from the *cycle basis* of $\Gamma_i$) are not catenated, one cannot immediately ensure the connectedness of HRN, although the author is not aware of any example. If HRN is a connected *crystallographic net*, we regard its automorphism group Aut(HRN) as *the maximal symmetry of an interpenetration pattern*. Now let us suppose that maximal symmetry $G_{max}$ is known for a system of interpenetrating nets $\Gamma$. The symmetry of the respective HRN($\Gamma$) is necessarily compatible with it. However, the intrinsic symmetry of HRN($\Gamma$) can be higher, as we shall see in the following. In general, we can write $G_{max}(\Gamma) \leq$ Aut[HRN($\Gamma$)]. For many interesting examples (and especially if all strong rings are catenated) $G_{max}(\Gamma)$ coincides with Aut[HRN($\Gamma$)]. Particularly, this holds for most frequent patterns **pcu-c**, **dia-c** and **srs-c** which serve as the labyrinths of *P*, *D* and *G* minimal surfaces, respectively:

HRN(**pcu-c**) = **nbo**, Aut(**nbo**) = $G_{max}$(**pcu-c**) = $Im\overline{3}m$;

HRN(**dia-c**) = **hxg**, Aut(**hxg**) = $G_{max}$(**dia-c**) = $Pn\overline{3}m$;

Aut[HRN(**srs-c**)] = $G_{max}$(**srs-c**) = $Ia\overline{3}d$.

Note that maximal symmetry of the HRN constructed for three-periodic labyrinths of three-periodic surfaces[9] could be also a tool to determine the inherent symmetry of the surfaces. Our work in this direction is in progress.

From the analysis of HRNs one can directly verify that '*normal*' (**dia-c5**) and '*abnormal*' (**dia-c5***) patterns of 5-fold interpenetrating **dia** nets are topologically distinct. The automorphism groups of their HRNs confirm that maximal symmetry of the former is $G_{max} = I4_1/amd$, $H_{max} = I4_1/amd$ (**a**, **b**, 5**c**) [in agreement with Bonneau and O'Keeffe, 2015] whereas the latter is indeed characterized by $G_{max} = I4_1/a$, $H_{max} = I4_1/a$ (2**a**–**b**, **a**+2**b**, **c**) as was deduced purely on the basis of group–supergroup relations. There are two ways to combine a pair of **dia-c5*** patterns together, either by a translation along [001] or a 2-fold rotation along [100]. The two possibilities are realized by space groups $P4_2/n$ and $I4_1/acd$, respectively. The first yields an arrangement **dia-c10*** with $G_{max} = P4_2/n$, $H_{max} = I4_1/a$ (3**a**–**b**, **a**+3**b**, 2**c**) (interpenetration class Ia), while the second gives rise to a different pattern **dia-c10\*\*** with $G_{max} = I4_1/acd$, $H_{max} = I4_1/a$ (2**a**–**b**, **a**+2**b**, **c**) (interpenetration class IIIa). To the knowledge of the author, the **dia-c10\*** pattern has been found among coordination polymers at least twice in the last years (Guo *et al.*, 2015; Tseng *et al.*, 2015), while **dia-c10\*\*** has never been observed to date. Furthermore, by carefully removing one net out of the **dia-c5\*** pattern, one generates a system of 4-fold interpenetrated *edge-transitive* **dia** nets [**dia-c4\***, $G_{max} = I4_1/acd$, $H_{max} = Fdd2$ (**a**–**b**, **a**+**b**, **c**)] where four nets are related by a $4_1$ screw axis (interpenetration class IIa, *cf*. Fig. 9 in Blatov *et al.*, 2004). All interpenetration patterns of **dia** nets discussed here can be constructed from individual nets in *ideal* geometries (see supporting information).

---

[9] This is not just an unnecessary word repetition because three-periodic surfaces may have layer-like, rod-like or even polyhedral 'labyrinths' (Koch, 2000*b*, 2001).

In practice it is convenient to calculate Aut[HRN(Γ)] using the *Gavrog Systre* code (Delgado-Friedrichs & O'Keeffe, 2003; http://gavrog.org/) and afterwards to look for the intersection group(s) $K$=Aut($Γ_i$)∩Aut[HRN(Γ)]. If the index of $K$ in Aut[HRN(Γ)] equals $n$ (*i.e.*, the number of interpenetrated nets), then group $K$ already corresponds to $H_{max}$ that maps a net $Γ_i$ onto itself in the set Γ with symmetry $G_{max}$ = Aut[HRN(Γ)]. In other words, to find $H_{max}$, one has to examine all subgroups $\widetilde{H}$ of $G_{max}$ = Aut[HRN(Γ)] with index $n$ and to choose it among those which have a subgroup relation to Aut($Γ_i$). If, however, $G_{max}(Γ) <$ Aut[HRN(Γ)], it is convenient first to search for the supergroups of Aut($Γ_i$) with index $n$ which have a subgroup relation to Aut[HRN(Γ)]. Note that in any case $H_{max} \leq$ Aut($Γ_i$). If supergroup search for Aut($Γ_i$) is not successful [or does not make sense if Aut[HRN(Γ)] ≤ Aut($Γ_i$)], it has to be performed for subgroups of Aut($Γ_i$).

*Example*: Koch *et al*. (2006) derived an interpenetrating sphere packing $t[4/4/t3]^2$ that is built up from the two distorted gismondine (**gis**) networks corresponding to $G = I4_1/acd$, $H = I4_1/a$. Let us determine the maximal symmetry of this pattern. The calculation using ToposPro (Blatov *et al*., 2014) yields that HRN($t[4/4/t3]^2$) = **hxg**. From the above discussion it follows that Aut(**hxg**) = $Pn\overline{3}m$. Aut(**gis**) = $I4_1/amd$, Aut(**hxg**)∩Aut(**gis**) = $I4_1/amd$. The only supergroup of $I4_1/amd$ with index 2 is $P4_2/nnm$ (that is in turn a subgroup of $Pn\overline{3}m$ with index 3). As a result, we obtain $G_{max}(t[4/4/t3]^2) = P4_2/nnm$, $H_{max} = I4_1/amd$. The embeddings with different symmetries ($P4_2/nnm$ vs. $I4_1/acd$) are compared in Fig. 2. In this example the group $G_{max}$ is uniquely defined that may not be always the case, as could be learnt from the next section.

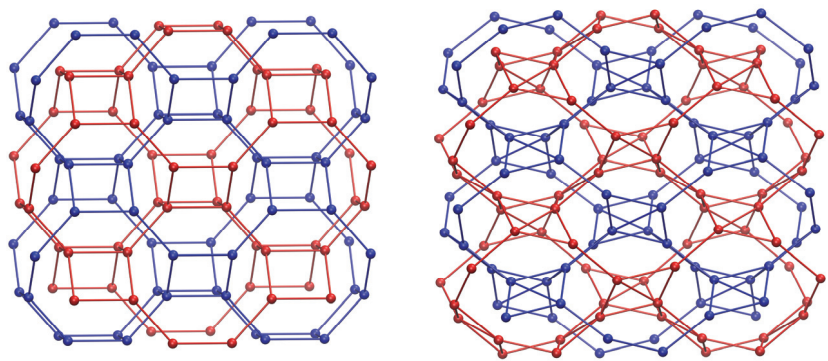

**Figure 2.** 2-fold interpenetrating **gis** nets in two distinct embeddings: in the maximal symmetry ($P4_2/nnm$, left) and as an interpenetrating sphere packing $t[4/4/t3]^2$ (symmetry $I4_1/acd$, right).

It is also helpful to consider the automorphism group of a *complete ring net* (CRN) [Baburin & Blatov, 2007; Alexandrov *et al.*, 2012] where two kinds of edges – the ones which stand for the Hopf links between rings and the ones which correspond to common edges of rings in the net $\Gamma_i$ – are both treated on equal footing. Following a suggestion from a referee, we shall reserve a term an *extended ring net* (ERN) for the CRN with both kinds of edges as we defined it here. The reason to introduce the ERN is that it becomes connected[10] (in contrast to HRN) for polycatenated systems (see Fig. 9 in Alexandrov *et al.*, 2012). Bonneau and O'Keeffe (2015) described two *inclined* polycatenation patterns **sql-c3\*\*** and **sql-c6** with symmetry $R3$ and $R3c$, respectively. The corresponding automorphism groups Aut(ERN) calculated using *Gavrog Systre* are, however, $R32$ and $Pm\bar{3}n$, which gave rise to the following embeddings (both patterns are actually *edge-transitive*):

**sql-c3\*\***, $R32$, $a$=1.0000, α=90°. Node at $3e$ (.2) 0.08839, 1/2, 0.91161. Link to 0.08839, 1/2, 1.91161.
**sql-c6**, $Pm\bar{3}n$, $a$=1.0000. Node at $6d$ ($\bar{4}m.2$) 0, 1/4, 1/2. Link to 0, 1/4, 3/2.

Finally, we would like to comment on the interpenetration and *parallel* polycatenation of 2-periodic layers. Although our approach based on the automorphism group of the HRN (or ERN) applies to such cases as well, it appears not to be very useful in practice. A space group of the corresponding HRN (or ERN) does not provide a recipe how nets should be undulated to allow for a reasonable embedding.

## 4. Working examples
### 4.1 Interpenetration patterns of two-fold *vertex-transitive* utp nets

In this section we systematically derive interpenetration patterns of two vertex-transitive **utp** nets. The **utp** graph was originally proposed by Wells as the (10,3)-*d* net (Wells, 1956, 1977) and was also found in the enumeration of sphere packings by Koch & Fischer (1995) under the name 3/10/*o*1. The maximal symmetry of **utp** is achieved in *Pnna* with the vertices in the general position 8*e*. From the *International Tables for Crystallography*, V. A1 (2004) it follows that there exist three supergroups of *Pnna* with index 2 which do not contain mirror planes, namely, *Ccce*, *Pcca* and *Pban*. These supergroups define three possibilities for a pair of (equivalent) **utp** nets to interpenetrate (Fig. 3). All three patterns belong to interpenetration class Ia, *i.e.*, individual nets can be mapped onto each other by translations (Blatov *et al.*, 2004). The arrangements with symmetry *Ccce* – *Pnna* (**utp-c\***)[11] and *Pcca* – *Pnna* (**utp-c\*\***) show an interesting property: both share the same HRN (**hxg**) and, hence, represent the interpenetration pattern **dia-c**. However, since *Pnna* is the maximal symmetry of a (single) **utp** net, both variants (*Ccce* – *Pnna* and *Pcca* – *Pnna*) can be considered as highest symmetry embeddings of 2-fold interpenetrated **utp** nets with the catenation pattern **dia-c**. The question of whether both entanglements are ambiently isotopic or not, remains open, although the author tends to agree with a referee that they are not. This example nevertheless demonstrates that the highest symmetry compatible with a given *system of interpenetrating*

---

[10] The ERN is necessarily connected for interpenetrating two- or three-periodic nets. However, it turns out to be very complicated to work with in practice (especially regarding the computation of the automorphism group).
[11] The **utp-c\*** arrangement can be realized as an interpenetrating sphere packing $o[3/10/o1]^2$ (Sowa, 2009).

*nets* (in contrast to the maximal symmetry of an *interpenetration pattern*) may not be uniquely defined. From a crystallographic point of view, a distinction between **utp-c*** and **utp-c**** should be made based on the arrangement of screws in a similar way as it was done by Fischer (1976) for interpenetrating sphere packings of type 3/3/*c*1 (**srs-a**). The **utp** net contains two kinds of (topological) 4-fold screws (Wells, 1977, p. 39) running along [010] (referred to the unit cell of *Pnna*), with four and eight vertices per translation period, respectively (*cf*. Fig. 3, bottom right). By comparing the locations of the screws, we see that in **utp-c*** the axes of different kinds of screws with opposite handedness coincide while in **utp-c**** different screws with the same handedness are co-axial (Fig. 3).

The **utp-c**** pattern (symmetry *Pban – Pnna*) is characterized by the same HRN as the pair of diamondoid networks in the structure of a coordination polymer [Ni(1,3-di(4-pyridyl)propane)(5-nitroisophthalato)($H_2O$)] reported by Xiao *et al*. (2005) and discussed in detail by Alexandrov *et al*. (2012).

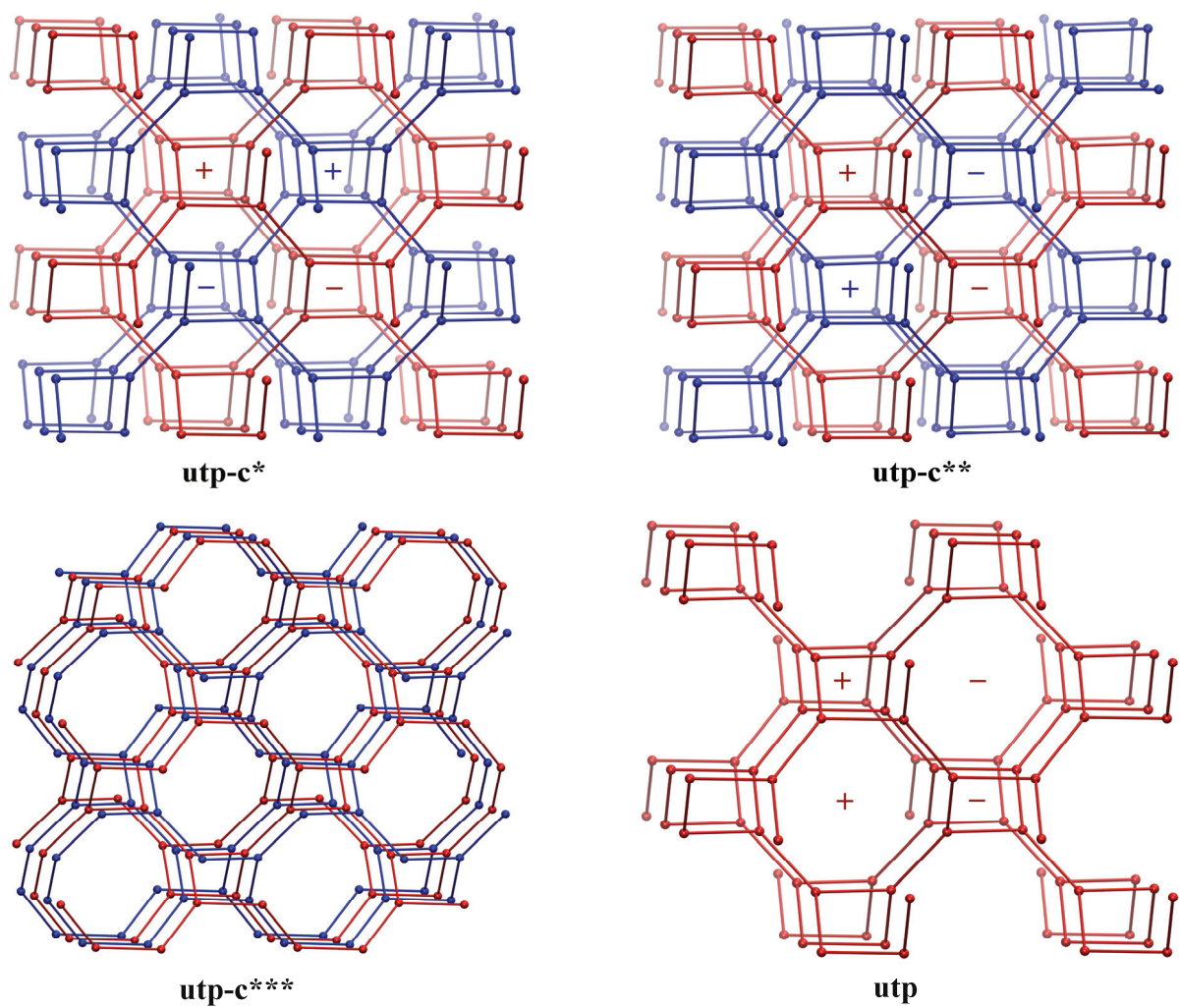

**Figure 3.** Interpenetration patterns of the **utp** net. Plus or minus signs refer to both location and handedness of the screws with the respective colours. In **utp-c**** screws of the same kind and handedness are co-axial and therefore not marked.

As the **utp** net is *bipartite*, it is of interest to analyze symmetries compatible with 2-fold interpenetrated copies of corresponding *binary* (two-colored) nets (**utp-b** following the nomenclature of

M. O'Keeffe, see *e.g.* Delgado-Friedrichs *et al.*, 2003), *i.e.* those where only vertices of two different colors are adjacent. The space group of **utp-b** is $Pna2_1$. Symmetry properties of 2-fold interpenetrating **utp** nets are summarized in Table 1. In contrast to connected graphs, a *bipartition* of a disconnected graph is not unique (Asratian, Denley & Häggkvist, 1998, Ch. 2); for this reason there are two symmetry alternatives for each pair of interpenetrated **utp-b** nets (Table 1). The vertices with unlike colors can be associated to different building blocks in a crystal structure, for example, metal connectors and ligands. One of the referees paid my attention to a family of coordination polymers with a general formula [M(4-pyrdpm)$_3$AgX] (M=Co, Fe, Ga, 4-pyrdpm = 5-(4-pyridyl)-4,6-dipyrrinato, X=$BF_4^-$, $PF_6^-$ *etc.*) which are based on tris(dipyrrinato)metalloligands and Ag$^+$ salts and crystallize in a space group *Pbcn* (Cohen *et al.*, 2006, 2007). In terms of topology, these structures can be described as 2-fold interpenetrating **utp-b** nets, with alternating 3-coordinated vertices representing Ag$^+$ cations and [M(4-pyrdpm)$_3$] metalloligands, respectively. Since the symmetry corresponds to the group–subgroup pair *Pbcn* – $Pna2_1$, the structures belong to the **utp-b-c\*** pattern (Table 1). Furthermore, there is another coordination polymer {[Zn(tipa)Cl]·NO$_3$·2DMF} (tipa=tris(4-imidazolylphenyl)amine), DMF = *N*,*N*-dimethylformamide) with a space group *Pnna* (Yuan *et al.*, 2014) that comprises 2-fold interpenetrating **utp-b** nets where Zn$^{2+}$ cations and tipa-ligands play the roles of alternating 3-coordinated nodes. In this case the symmetry is *Pnna* – $Pna2_1$ and we identify the **utp-b-c\*\*\*** pattern (Table 1). After iodine uptake the structure undergoes a transition to a phase {[Zn$_2$(tipa)$_2$Cl$_2$]·2I$_3$·2DMF} with a different framework topology, that of a *binodal* **cfc**-3,3-*Pbcn* net (the nomenclature of Blatov & Proserpio, 2009) but still remains 2-fold interpenetrated (the symmetry becomes *Pbcn* – $Pca2_1$). Topological changes were overlooked by the authors who assumed that after the transition "the coordination network as well as the topology were maintained" (Yuan *et al.*, 2014, p. 10095).[12]

**Table 1.** Symmetry properties of 2-fold interpenetrating **utp** nets[*]

| Pattern | Maximal symmetry *(group–subgroup)* | Interpen. class | Pattern | Possible symmetry *(group–subgroup)* | Interpen. class |
|---|---|---|---|---|---|
| **utp-c\*** | *Ccce* – *Pnna* | Ia | **utp-b-c\*** | ***Pbcn* – $Pna2_1$** <br> *Aea*2 – $Pna2_1$ | IIa <br> Ia |
| **utp-c\*\*** | *Pcca* – *Pnna* | Ia | **utp-b-c\*\*** | *Pccn* – $Pna2_1$ <br> $Pca2_1$ – $Pna2_1$ | IIa <br> Ia |
| **utp-c\*\*\*** | *Pban* – *Pnna* | Ia | **utp-b-c\*\*\*** | ***Pnna* – $Pna2_1$** <br> *Pba*2 – $Pna2_1$ | IIa <br> Ia |

[*] The symmetries shown in bold correspond to the observed crystal structures with the same interpenetration pattern (centrosymmetric intergrowths are preferred).

According to a private communication from a referee, there also exists a crystal structure [Cu$_2$I$_2$L$_2$] (L = *N*, *N*, *N*´, *N*´-tetrakis(4-pyridyl)-1,4-phenylenediamine, Tang *et al.*, 2013) which is related to the **utp-c\*\*** pattern. The symmetry is $P2_1/n$ – $P2_1/c$ (2**a**, **b**, **a**+**c**). Here the **utp** topology can be assigned to frameworks if Cu$^+$ centers and tri-coordinated *branching points* of a tetratopic linker (*cf.* O'Keeffe &

---
[12] The nets **utp** and **cfc**-3,3-*Pbcn* are non-isomorphic but, however, very similar to each other since both can be viewed as *isohedral* tilings (Blatov *et al.*, 2007) by the same [10$^4$] cages (just as the **ths** net) and share a common *dual* net (diamond).

Yaghi, 2012; Li *et al*., 2014) are treated as the nodes of the underlying net. This case stands apart from the examples discussed above where the nodes always represent entire chemical species.

**4.2 Novel *two-fold* interpenetrated *vertex-transitive* dia and pcu nets**

To the knowledge of the author, the only interpenetration patterns of 2-fold intergrown **pcu** and **dia** – if individual nets remain in their ideal configurations – are accordingly **dia-c** and **pcu-c**. Recently, Zaworotko *et al*. (2015) discovered a coordination polymer [Ni(1,2-bis(4-pyridyl)acetylene)$_2$(Cr$_2$O$_7$)] where two **pcu** nets interpenetrate in a different way that corresponds to the group-subgroup pair $I$222 – $I$112 (interpenetration class IIa). However, in this arrangement **pcu** nets need to deviate from their most symmetric configuration to allow for this unusual kind of catenation. The maximal symmetry of this pattern is $Ibam$ (the space group of the individual net is $I$112/$m$), *i.e.* it is not intrinsically chiral, as can be shown by the analysis of its HRN (see Supporting information).

The diamond net and the primitive cubic lattice have high inherent symmetries, namely, $Fd\bar{3}m$ and $Pm\bar{3}m$, respectively. To keep them in the ideal conformations and at the same time to replicate by a certain space group avoiding edge crossings is not easy if one does not go systematically through all subgroups of **dia** and **pcu**. However, to provide an example, it is reasonable to replicate both (intrinsically) cubic nets by a space group that has no group–subgroup relation to their inherent symmetries, *e.g.* a hexagonal space group in this case. It is known (Koch & Sowa, 2005) that owing to subgroup relations both *ideal* **dia** and **pcu** nets can occur in space group $P3_121$ with all the vertices in the general position (6$c$). By 'embedding' a net (either **dia** or **pcu**) formally described in $P3_121$ 6$c$ into *e.g.* its supergroup $P6_122$, we arrive at the two *unprecedented* interpenetration patterns which we label as **dia-c\*** and **pcu-c\*** (both of class IIa, Fig. 4). The automorphism groups of their HRNs prove that the inherent symmetries of both patterns are accordingly $P6_122$ (**dia-c\***) and $P6_222$ (**pcu-c\***) (see Supporting information).

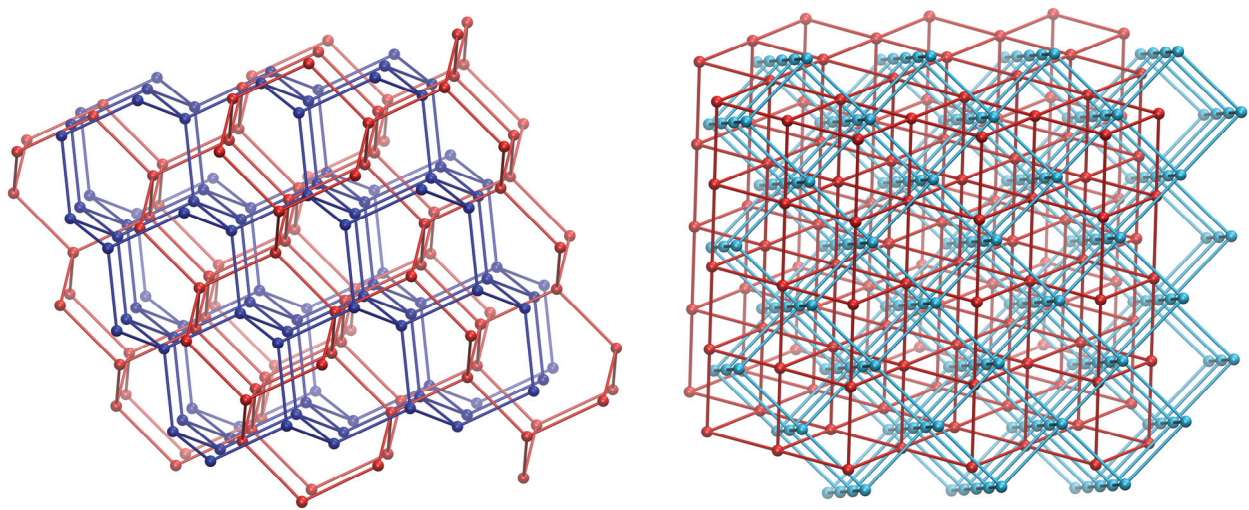

**Figure 4**. Novel *hexagonal* patterns **dia-c\*** (left) and **pcu-c\*** (right).

## 4.3 *Enantiomorphic versus chiral* interpenetrating quartz nets

Now let us think of how two quartz (**qtz**) nets, both in their *ideal* configurations but with opposite handedness, could be combined together. Interpenetration patterns of **qtz** nets have included only chiral examples until recently (Baburin *et al.*, 2005; Alexandrov *et al.*, 2011). Here we explain the procedure to derive interpenetration pattern(s) of enantiomorphic **qtz** nets. First, we observe that none of the vertex-transitive subgroups of the **qtz** net ($P6_222$, $P6_422$ (2**c**), $P6_122$ (2**c**), $P6_522$ (4**c**), $P6_1$ (2**c**), $P6_2$, $P3_221$, $P3_212$, $P3_121$ (2**c**), $P3_112$ (2**c**), $P3_2$) has a supergroup which contains symmetry operations of the second kind (*i.e.*, improper rotations). This means that there does not exist a space group that acts transitively on the vertices of two (or more) enantiomorphic **qtz** nets (in fact, this statement applies to any vertex-transitive net having the $P6_222$ group as maximal symmetry). As a result, we are forced to go through vertex-2-transitive subgroups. The search for such subgroups was facilitated by using the ToposPro package (Blatov & Proserpio, 2009; Blatov *et al.*, 2014). Among vertex-2-transitive subgroups of the **qtz** net, the $C222$ group stands out because the parent 222 ($D_2$) point-group symmetry is still kept at one of the nodes. The $C222$ group has only one *achiral* orthorhombic supergroup without mirror planes[13], namely the group $Ccce$. By 'embedding' the **qtz** net (referred to $C222$) into the $Ccce$ group, we generate the desired centrosymmetric pattern **qtz-c\*** (Fig. 5). Other lower-symmetry alternatives (group–subgroup pairs *e.g.* $Pccn$ – $P2_12_12$, $Pbcn$ – $P2_12_12$ or $C2/c$ – $C2$) correspond topologically to the same entanglement (**qtz-c\***) as the pair $Ccce$ – $C222$ (interpenetration class IIa). Note that 2-fold enantiomorphic quartz nets have been recently observed in the monoclinic ($C2/c$) structure of a coordination polymer with a framework composition $Zn(bmzbc)_2$ [Hbmzbc=4-(benzimidazole-1-yl)benzoic acid] (Wang *et al.*, 2015).

The analysis of catenated rings in **qtz-c\*** using ToposPro has revealed only Hopf links, in contrast to the chiral **qtz-c** pattern (Fig. 5) which also contains doubly-interlaced 8-rings [so-called *Solomon links*, Nierengarten *et al.*, 1994] (*cf.* Fig. 10 in Bonneau and O'Keeffe, 2015). The absence of intrinsically chiral Solomon links in the intergrowth of enantiomorphic quartz nets is curious because one could expect equal amounts of 'left' and 'right' links to coexist in the racemate. However, the **qtz-c** pattern is not the only variant for a pair of homochiral **qtz** nets to be intertwined. For example, two **qtz** nets with symmetry $P6_222$ – $P3_212$ (**qtz-c\*\*** pattern, interpenetration class IIa, Fig. 5) can interpenetrate and also avoid multiply-interlaced rings. Surprisingly, the quartz-like labyrinths of $HS2$ minimal surface (Fischer & Koch, 1989) show the same symmetry ($P6_222$ – $P3_212$) and also interpenetrate without multiply-interlaced rings but in a way different from **qtz-c\*\***.

Chiral intergrowths of *n*-fold **qtz** nets with symmetry $P6_222$ or $P6_422$ discussed by Bonneau and O'Keeffe (2015) can be similarly described in $C222$ and replicated by the $Ccce$ group. In total, the whole arrangement will contain $2n$ enantiomorphic **qtz** nets. Such patterns necessarily belong to interpenetration class III. The simplest pattern (**qtz-c4***) comprising two pairs of **qtz** nets with opposite handedness is given in the supporting information.

---

[13] It also contains tetragonal supergroups ($P\bar{4}c2$, $P\bar{4}b2$, $P\bar{4}n2$ *etc.*) but due to metrical problems the transformation from orthorhombic to tetragonal basis would necessarily induce distortions of the individual nets.

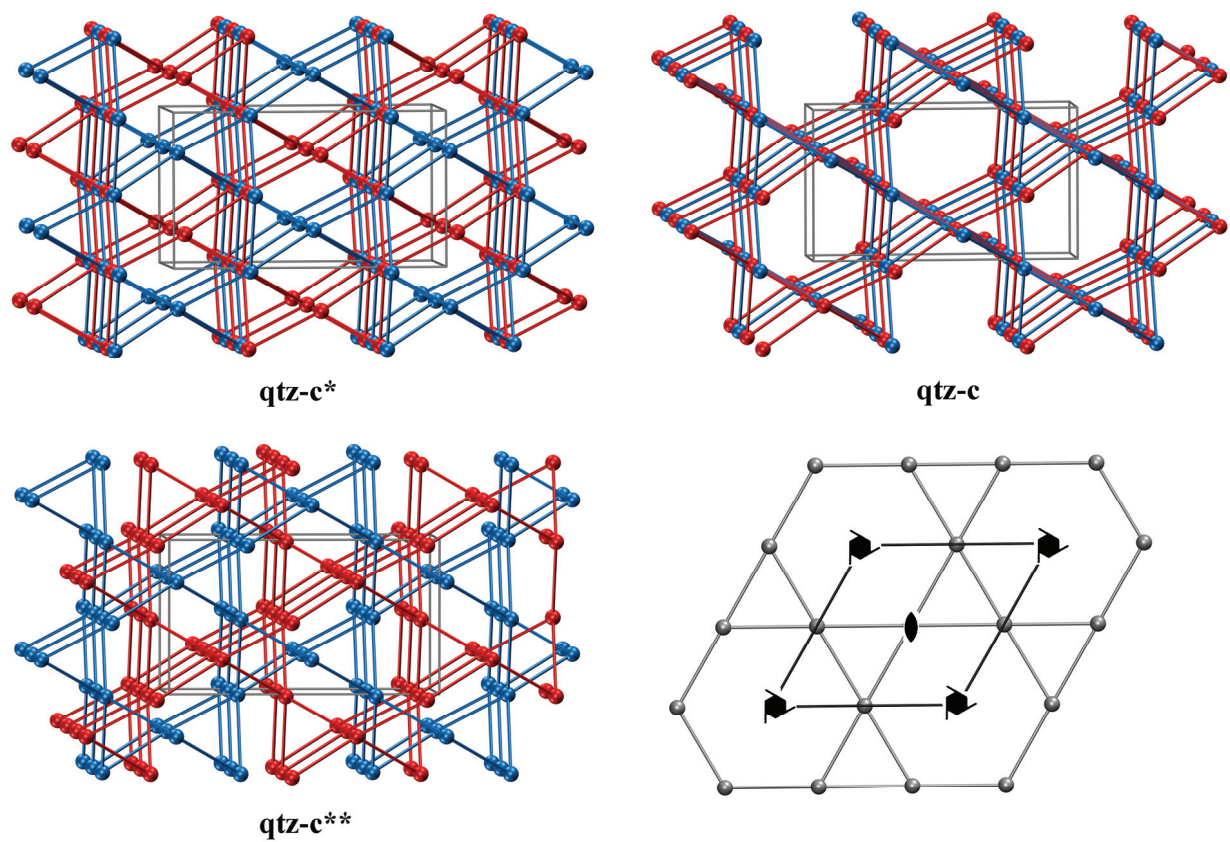

**Figure 5**. 2-fold interpenetrating **qtz** nets. Notice that in **qtz-c*** the $6_2$ screws of the blue net coincide with the 2-fold rotation axes of the red net (and *vice versa*) that is an evidence for orthorhombic symmetry ($6_2 \cap 2 = 2$). Two hexagonal patterns (**qtz-c** and **qtz-c****) are drawn in orthohexagonal setting for a better comparison with **qtz-c***. At the bottom right we illustrate the location of some screw and rotation axes along [001] in the conventional unit cell of quartz.

## 5. Conclusion

In this paper we presented a general group-theoretical approach to systematically construct interpenetration patterns based on group–supergroup relations. Given a net embedding with a space group *H*, it can be replicated by its supergroup *G* (of finite index) under the condition that site-symmetries of vertices and edges are the same in both *H* and *G*. We showed that edge crossings are unavoidable whenever interpenetrating nets are mapped onto each other by mirror reflections. More generally, edge crossings are caused by any symmetry element of finite order from *G* \ *H* that intersects vertices and/or edges of the nets. This property restricts the number of supergroups to be considered for the generation of interpenetrating nets. We proposed a systematic procedure to determine maximal symmetries of interpenetration patterns by using the automorphism group of a *Hopf ring net* (or alternatively, an *extended ring net*) and its subgroup relations to possible symmetries of individual nets. Following our approach, we completely derived intergrowths of two vertex-transitive **utp** nets. We discovered unprecedented arrangements of 2-fold **dia** and **pcu** where the nets retain their ideal configurations. Furthermore, we found the highest symmetry embedding for a pair of enantiomorphic quartz (**qtz**) nets and showed that the corresponding space group (*Ccce*) is vertex-2-transitive.

**Acknowledgements**

I am deeply indebted to Prof. Vladimir I. Trofimov (Ekaterinburg, Russia) for valuable comments on the action of groups on graphs. I would like to thank Prof. Davide Proserpio (Milano, Italy) for many fruitful discussions in April 2015 during his visit to Dresden and for suggesting to me to think about enantiomorphic quartz nets. I thank M. Sc. Sergei Bronnikov (Samara, Russia) for helpful conversations. I am very grateful to one of the referees for constructive comments and suggestions which significantly improved an early version of the paper. Financial support from the EU within the "Graphene Flagship" is acknowledged.

**Supporting information**

Supporting information contains the coordinates of all the structures (cgd-format) mentioned in the paper.